\documentclass[11pt,a4paper]{article}
\usepackage[left=15mm, top=10mm, right=15mm, bottom=20mm, nohead=true, nofoot=true]{geometry}

\usepackage[utf8]{inputenc}
\usepackage{authblk}
\usepackage{indentfirst}
\usepackage{misccorr}
\usepackage{graphicx}
\usepackage{amsmath}
\usepackage{amssymb}
\usepackage{multicol}
\usepackage{bm}
\usepackage{longtable}
\usepackage{pdflscape}
\usepackage{epstopdf}
\usepackage{multirow}
\usepackage{adjustbox}
\usepackage{amsfonts}
\usepackage{ifthen}
\usepackage{fancyhdr}
\usepackage{setspace}
\usepackage{xcolor}
\usepackage[round,authoryear,comma,sort&compress]{natbib}
\usepackage[toc,title,page]{appendix}
\usepackage[hidelinks]{hyperref}
\usepackage{url}
\usepackage{booktabs}
\newcommand{\tablecomments}[1]{\par\vspace{2pt}\noindent #1\par}

\hypersetup{
    colorlinks=true,
    linkcolor=green,
    citecolor=blue,
    filecolor=magenta,
    urlcolor=cyan,
    pdftitle={TRGB Distances to the Andromeda Dwarf Satellites from the INT Monitoring Survey},
    pdfpagemode=FullScreen,
}
\fancypagestyle{plain}{\chead{\texttt{\textcolor{brown}{Communications of BAO, Vol. 73, Issue 2, 2026, pp. ???-???}}}}
\title{\textbf{TRGB Distances to the Andromeda Dwarf Satellites from the Isaac Newton Telescope Monitoring Survey}}
\author[1, 2]{Hedieh Abdollahi\thanks{hedieh.abdollahi@csfk.org}}

\affil[1]{\scriptsize Konkoly Observatory, HUN-REN Research Centre for Astronomy and Earth Sciences, MTA Centre of Excellence, Konkoly-Thege Mikl\'os \'ut 15-17, H-1121, Budapest, Hungary}
\affil[2]{\scriptsize School of Astronomy, Institute for Research in Fundamental Sciences (IPM), P.O. Box 1956836613, Tehran, Iran}

\begin{document}
\pagestyle{empty}
\newpage
\pagestyle{fancy}
\label{firstpage}
\date{}
\maketitle

\begin{abstract}
We measure tip of the red giant branch (TRGB) distances to 17 dwarf satellites of M31 from Sloan $i$-band photometry of the Isaac Newton Telescope (INT) monitoring survey of Local Group dwarf galaxies, and add the distances of And~I, And~VII and And~IX from earlier papers of the survey. We locate the tip with a Sobel edge-detection filter applied to the luminosity function of stars within two half-light radii of each galaxy. The distance moduli range from $23.38 \pm 0.06$\,mag (And~XVI, 474\,kpc) to $25.35 \pm 0.06$\,mag (And~XVIII, 1177\,kpc), and 15 of the 20 satellites lie between 650 and 850\,kpc. For most galaxies, our distance agrees with at least one published value. Compared with \citet{Conn2012}, our tip is brighter for 12 of the 18 galaxies in common, by up to 0.39\,mag, possibly because of crowding in the central regions. And~XII lies at $844^{+40}_{-38}$\,kpc, about 60\,kpc behind M31, consistent with a first infall into the Local Group. And~XXII lies at $753^{+28}_{-27}$\,kpc, closer than all earlier estimates, and about 59\,kpc from M33 in three dimensions (for an M33 distance of 794 kpc), consistent with an association with M33. And~XVIII lies about 400\,kpc behind M31, too far for a secure association with M31.
\end{abstract}

\emph{\textbf{Keywords:} stars: luminosity function, mass function --
galaxies: distances and redshifts --
galaxies: dwarf --
galaxies: stellar content --
Local Group}

\section{Introduction}

The three-dimensional positions of the M31 satellites show how the satellite system is arranged and which galaxies belong to M31 or to M33. Their positions on the sky are known precisely, so the main uncertainty is their distance along the line of sight. For several of the fainter satellites, published distance moduli differ by a few tenths of a magnitude \citep{Conn2012,McConnachie2012}. At the distance of M31 ($\sim780$\,kpc), a difference of 0.3\,mag moves a satellite by about 110\,kpc along the line of sight, enough to change whether it appears to lie near M31, near M33, or far from both. Distance alone cannot establish membership, but it is the first constraint, to be combined with kinematics. Reliable distances are also needed to compare the stellar populations of M31 and Milky Way satellites on the same basis.

The tip of the red giant branch (TRGB) is a standard distance indicator for resolved galaxies. It shows up as a sharp break in the luminosity function of the red giant branch, and in red optical bands its absolute magnitude varies little with age and metallicity in old, metal-poor populations \citep{Lee93}. \citet{Conn2012} measured TRGB distances to most M31 satellites with a Bayesian method built for sparsely populated systems, and \citet{2022ApJ...938..101S} have since provided homogeneous RR~Lyrae distances for most of them. Faint dwarfs remain difficult: few stars lie near the tip, and the result depends on the photometric depth, the extinction correction and the method used to find the edge. Published distances therefore differ partly because the data and analyses differ \citep{Mahani23}. Measuring all satellites from one data set with one method reduces this source of scatter.

Our data come from the Isaac Newton Telescope (INT) monitoring survey of Local Group dwarf galaxies, which was designed to find long-period variable stars (LPVs) and study the evolved stellar populations of these systems \citep{2020ApJ...894..135S}. LPVs trace their parent populations, because their luminosities, pulsation properties and evolutionary timescales are linked to the properties of their progenitor stars \citep{Mahani_Farsi}. In M33, near-infrared LPV monitoring has been used to reconstruct the star-formation history of the centre and the disc \citep{Javadi11,2011MNRAS.414.3394J,2017MNRAS.464.2103J} and to study dust production by evolved stars \citep{Javadi13}. Similar work has been done for NGC~147 and NGC~185 \citep{Mahani25} and for NGC~6822 \citep{Rostami25}, using SED-fitting tools such as SEDust \citep{Mahani_Bao}. The same INT images also give deep, uniform $i$-band photometry of the red giant branch, which we use here to measure TRGB distances.

The variable-star catalogue of the survey is published in \citet{Abdollahi2026,2026yCat..20000069A}. Here, we present the TRGB distances of the observed M31 satellites, compare them with earlier estimates, and discuss three galaxies, And~XII, And~XXII and And~XVIII, whose distances affect which systems belong to M31 or M33.

\section{Observations and data}
\label{sec:data}

The observations were obtained with the Wide Field Camera (WFC) on the 2.5\,m Isaac Newton Telescope at the Observatorio del Roque de los Muchachos, La Palma, between June 2015 and February 2018. The WFC consists of four $2048 \times 4096$ CCDs, with a pixel scale of 0.33\,arcsec\,pixel$^{-1}$. Each galaxy was observed at multiple epochs in the Sloan $i$ and Harris $V$ filters. The survey strategy and the number of epochs for each galaxy are described in \citet{2020ApJ...894..135S}.
Of the 24 M31 satellites in the survey sample, four (M110, M32, Pisces~I and the Pegasus dwarf irregular, DDO~216) were not observed often enough for the variability analysis. And~I, And~VII and And~IX were analysed in earlier papers of the survey \citep{2021ApJ...923..164S,2021ApJ...910..127N,Abdollahi2023}. We include their distances for completeness, so the final sample has 20 galaxies.

The images were reduced and aligned using the \textsc{theli} pipeline \citep{Erben05}. We performed the photometry with \textsc{daophot/allstar} and \textsc{allframe} \citep{Stetson1987,Stetson94}, which are well suited to the crowded fields of these galaxies, and calibrated the photometry using standard stars from \citet{Landolt92}. For most galaxies, we used only CCD4, which covers approximately two half-light radii. For the more extended systems, such as And~II \citep{Abdollahi2024} and And~XIX, we also used the other CCDs. Artificial-star tests show that the photometry is 90--95 per cent complete down to $i \approx 22$\,mag, which is fainter than the TRGB in all targets except And~XVIII, and about 50 per cent complete at $i \approx 23$\,mag.

\section{Measuring the TRGB}
\label{sec:method}

The TRGB marks the onset of core helium burning in low-mass red giants. In the Cousins $I$ band, its absolute magnitude changes little with age and metallicity in old, metal-poor populations \citep{Lee93}, and the tip has also been calibrated in the SDSS $i$ band \citep{Bellazzini2008}. In the luminosity function, the tip appears as a sudden rise in star counts toward fainter magnitudes, so it can be found by edge detection.

\subsection{Edge detection}

\citet{Lee93} located the tip by applying a Sobel filter to the binned luminosity function. With $N_k$ the number of stars in bin $k$ ($k$ increasing toward fainter magnitudes), the three-point kernel $[-1,0,+1]$ gives
\begin{equation}
E_k = N_{k+1} - N_{k-1},
\label{eq:sobel3}
\end{equation}
and the tip is placed at the strongest positive response. This kernel is equivalent, up to a factor of two, to the $[-2,0,+2]$ form.

The result depends on the bin width and on where the bin edges fall \citep{Makarov06}. \citet{Sakai96} avoided binning by representing each star as a Gaussian with a width set by its photometric error, but too much smoothing can shift the measured tip, especially in noisy data \citep{Madore2023}. For our noisiest luminosity functions, we used the five-point kernel $[-2,-1,0,+1,+2]$ \citep{Abdollahi2026}, which spans more bins and suppresses spurious peaks.

\subsection{Implementation and distance}

\begin{figure}[htbp]
\centering

\begin{minipage}[t]{0.48\textwidth}
    \centering
    \includegraphics[width=\linewidth]{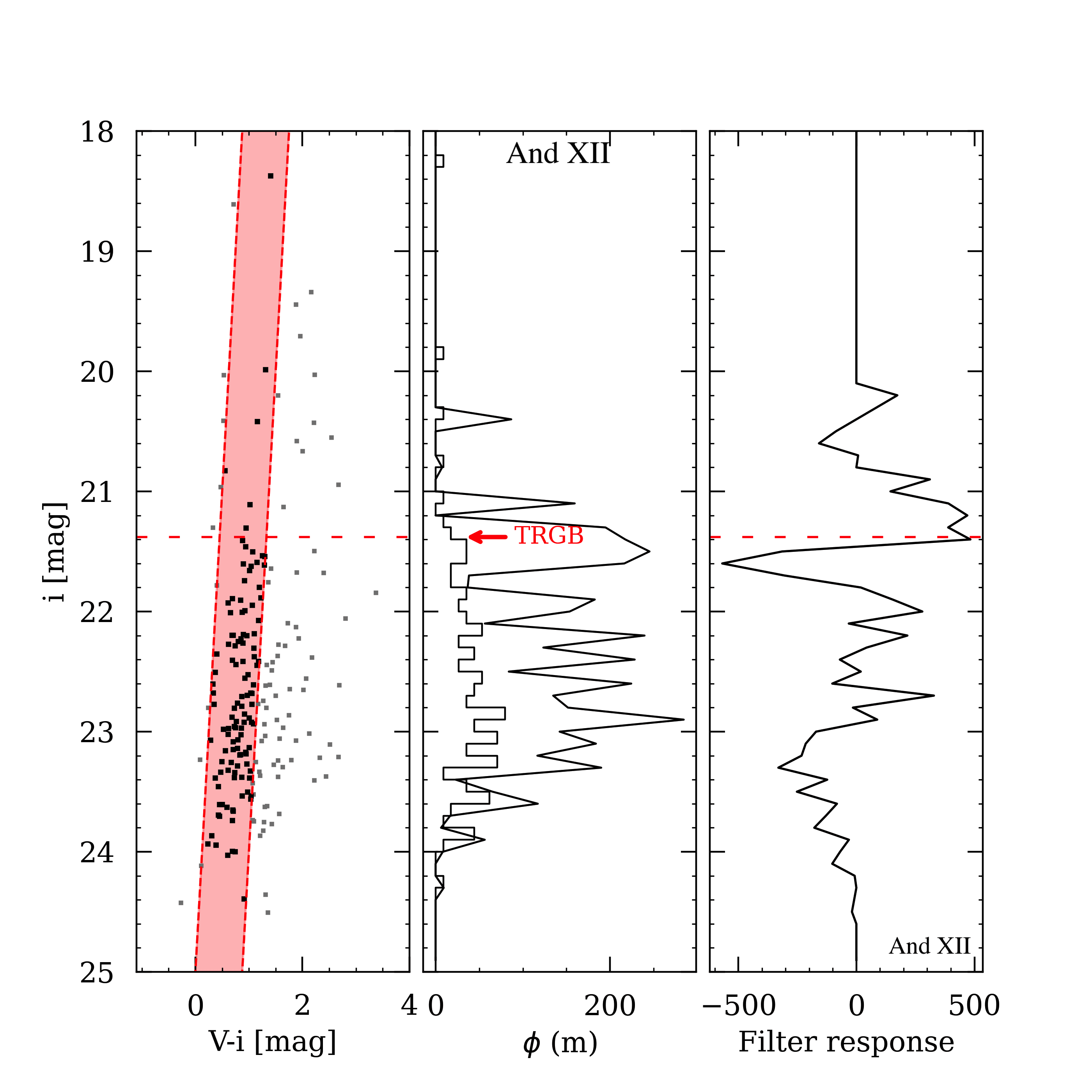}
    \caption{TRGB measurement for And~XII. Left: stars within
    two half-light radii; the shaded band is the RGB colour window. Middle: $i$-band luminosity function. Right: Sobel filter response; the red line marks the TRGB.}
    \label{fig:sobel}
\end{minipage}
\hfill
\begin{minipage}[t]{0.46\textwidth}
    \centering
    \includegraphics[width=\linewidth]{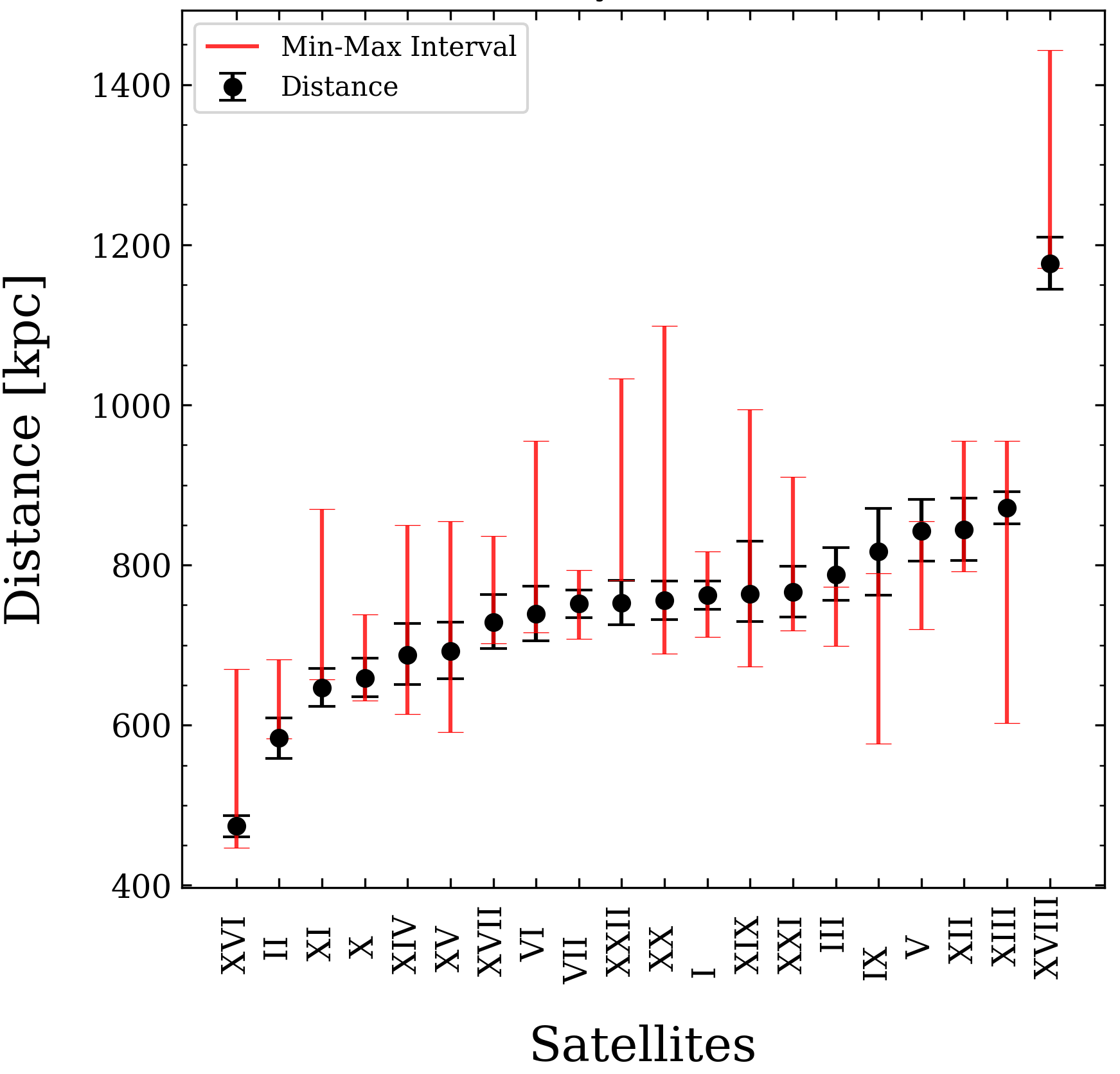}
    \caption{TRGB distances of the M31 satellites from this work
    (black circles) compared with the range of published values
    for each galaxy (red bars; values in Table~\ref{table:TRGB}).}
    \label{fig:distances}
\end{minipage}

\end{figure}

For each galaxy, we selected stars within two half-light radii of its centre \citep{Abdollahi2026,Abdollahi2025} and within a colour window around the red giant branch (shaded band in the left panel of Figure~\ref{fig:sobel}). We built the $i$-band luminosity function with 0.05\,mag bins, applied the Sobel filter, and took the strongest response as the tip, after checking the luminosity function for competing peaks. Figure~\ref{fig:sobel} shows the result for And~XII. Spurious detections become more likely when few RGB stars lie near the tip and photometric errors are large \citep{Madore2023}, which affects the faintest satellites in our sample. \citet{2022ApJ...938..101S} also found that TRGB distances to galaxies fainter than $M_V\approx-9.5$ can be biased by up to 35 per cent.

We then computed the distance modulus as
\begin{equation}
\mu = m_{\rm TRGB} - A_i - M_{\rm TRGB},
\label{eq:mu}
\end{equation}
where $M_{\rm TRGB}$ is the absolute tip magnitude in the SDSS $i$ band, which depends only weakly on metallicity for old, metal-poor populations \citep{Lee93,Bellazzini2008}. The adopted values are listed in Table~\ref{table:TRGB}. The foreground extinction is $A_i = 1.698\,E(B-V)$, with $E(B-V)$ from \citet{Schlegel1998} and the coefficient of \citet{Schlafly2011} for $R_V=3.1$. The uncertainties on $\mu$ include only the tip measurement, not the systematic uncertainties of $M_{\rm TRGB}$ and of the extinction correction.

\begin{table*}
\centering
\caption{TRGB magnitudes and distances of the targets, compared with published values.}
\small 
\setlength{\tabcolsep}{2.1pt}
\label{table:TRGB}
\centering
\begin{tabular}{llllllll}
\hline\hline
 \noalign{\smallskip}
{Galaxy} &
TRGB$_{\,this\,work}$ (mag) &
$\mu_{\,this\,work}$ (mag) &
TRGB$^a$ (mag)&
$\mu_{Min}$ (mag) &
$\mu_{Max}$ (mag)&
$M_{\rm TRGB}$ (mag)
&
A$_i$  (mag)\\
&  &Distance (kpc) &  & Distance (kpc) & Distance (kpc) &&\\

 \noalign{\smallskip}
\hline
 \noalign{\smallskip}
And\,I$^\star$      & $21.00\pm0.05$ & $24.41\pm0.05$ & $20.98\pm0.05$ &$24.31\pm0.05$ $^a$ & $24.56\pm0.09$ $^b$ & $-3.50$ & 0.090\\
            & & $762.08^{+17.75}_{-17.35}$& &$727.28\pm16.95$ & $816.58\pm 34.56$& &\\
            
And\,II     & $20.40\pm0.10$ & $23.83\pm0.10$ & $20.69\pm0.05$ &$23.83\pm0.42$ $^c$ & $24.17\pm0.12$ $^b$ & $-3.53$ & 0.104 \\
            & & $584.45^{+27.54}_{-26.31}$& &$583.96\pm54.69$ & $682.34\pm 38.77$& &\\

And\,III   & $21.14\pm0.09$ & $24.48\pm0.09$ & $20.98\pm0.07$ & $24.18\pm0.11$ $^d$ & $24.48\pm0.19$ $^b$ & $-3.43$ & 0.094\\
           & & $788.21^{+33.36}_{-32.00}$ & &$685.44\pm30.98$ & $787.17\pm68.41$ & & \\

And\,V     & $21.27\pm0.10$ & $24.63\pm0.10$ & $21.17\pm0.07$&$24.35\pm0.07$ $^a$ & $24.66\pm0.13$ $^e$ & $-3.57$ & 0.212\\
           & & $842.46^{+39.70}_{-37.92}$& & $741.31\pm24.29$ & $855.07\pm 52.75$ & &\\

And\,VI    & $21.01\pm0.10$ & $24.34\pm0.10$ & - & $24.37\pm0.09$ $^d$ & $24.94\pm0.09$ $^f$ & $-3.44$ & 0.107 \\
           & &$738.93^{+34.83}_{-33.26}$& & $748.17\pm31.66$ & $973.99\pm40.42$ & &\\

And\,VII$^\star$   & $21.30\pm0.05$ & $24.38\pm0.05$ & - &$24.25\pm0.25$ $^g$ & $24.50\pm0.10$ $^h$ & $-3.41$ & 0.329\\
           & & $751.62^{+17.51}_{-17.11}$& &$707.95\pm86.38$ & $794.33\pm 37.43$ & & \\

And\,IX$^\star$    & $21.20\pm0.15$ & $24.56\pm0.15$& $20.61\pm0.31$ &$23.89\pm0.31$ $^a$ & $24.48\pm0.20$ $^i$ & $-3.49$ & 0.128\\
           & &$816.58^{+58.40}_{-54.50}$& &$599.79\pm92.04$ & $787.05\pm 75.93$ & &\\

And\,X     &$20.77\pm0.08$ & $24.09\pm0.08$ & $20.95\pm0.28$& $24.01\pm0.07$ $^j$ & $24.23\pm0.10$ $^k$ & $-3.54$ & 0.219\\
           & & $659.19^{+24.74}_{-23.84}$& &$633.87\pm20.77$ & $701.46\pm33.05$ & &\\

And\,XI    & $20.75\pm0.08$& $24.05\pm0.08$ & $21.14\pm0.31$& $24.17\pm0.09$ $^d$ & $24.70\pm0.20$ $^l$ & $-3.44$ & 0.136\\
           & & $646.89^{+24.28}_{-23.40}$& & $682.34\pm28.87$ & $870.96\pm 84.03$ & & \\

And\,XII   & $21.38\pm0.10$ & $24.63\pm0.10$ & $21.63\pm0.34$&$24.44\pm0.12$ $^f$ & $24.84\pm0.34$ $^a$ & $-3.44$ & 0.187\\
           & & $843.90^{+39.77}_{-37.99}$& & $773.76\pm47.26$ & $928.97\pm 157.46$ & & \\

And\,XIII  & $21.40\pm0.05$ & $24.70\pm0.05$ & $21.30\pm0.07$ &$24.40\pm0.49$ $^a$ & $24.80\pm0.35$ $^m$ & $-3.44$ & 0.140\\
           & & $871.27^{+20.29}_{-19.83}$ & &$758.58\pm192.02$ & $912.01\pm 159.51$ & & \\

And\,XIV   & $20.85\pm0.12$ & $24.19\pm0.12$ & $21.18\pm0.56$ &$24.33\pm0.33$ $^n$ & $24.50\pm0.56$ $^a$ & $-3.44$
& 0.101\\
           & & $688.06^{+39.10}_{-37.00}$ & &$734.51\pm120.56$ & $794.33\pm 233.71$ &&\\
           
And\,XV    & $20.84\pm0.11$ & $24.21\pm0.11$ & $21.02\pm0.05$ &$23.98\pm0.26$ $^a$ & 24.66 $^o$ & $-3.45$ & 0.079\\
           & & $692.43^{+35.98}_{-34.20}$& &$625.17\pm 79.52$ & 855.07 &&\\

And\,XVI   & $20.05\pm0.06$ & $23.38\pm0.06$ & $20.27\pm0.08$ &$23.30\pm0.19$ $^a$ & $23.72\pm0.09$ $^p$ & $-3.44$ & 0.113\\
           & & $473.79^{+13.27}_{-12.91}$& &$457.09\pm 41.80$ & $554.62\pm23.47$ &&\\

And\,XVII  & $21.00\pm0.10$ & $24.31\pm0.10$ & $21.12\pm0.07$ &$24.31\pm0.11$ $^a$ & $24.50\pm0.10$ $^q$ & $-3.44$ & 0.126\\
           & & $728.67^{+34.34}_{-32.80}$ & &$727.78\pm37.82$ & $794.33\pm37.43$ &&\\
           
And\,XVIII & $22.09\pm0.06$ & $25.35\pm0.06$ & $22.20\pm0.08$ &$25.42\pm0.08$ $^a$ & $25.66\pm0.13$ $^r$ & $-3.44$ & 0.180\\
           & & $1176.74^{+32.97}_{-32.07}$& & $1213.39\pm45.54$ & $1355.19\pm 83.61$ &&\\
           
And\,XIX   & $21.08\pm0.10$ & $24.41\pm0.10$ & $21.24\pm0.09$ &$24.52\pm0.23$ $^s$ & $24.85\pm0.13$ $^r$ & $-3.44$ & 0.105\\
           & & $763.74^{+35.99}_{-34.38}$ & &$801.68\pm89.57$ & $933.25\pm 57.58$ &&\\
           
And\,XX    & $21.05\pm0.07$ & $24.39\pm0.07$ & $21.03\pm0.08$ &$24.35\pm0.16$ $^a$ & $24.52\pm0.74$ $^r$ & $-3.44$ & 0.100\\
           & & $755.62^{+24.75}_{-23.97}$ & & $741.31\pm56.68$ & $801.68\pm325.52$ &&\\
           
And\,XXI   & $21.14\pm0.09$ & $24.42\pm0.09$ & $21.19\pm0.07$&$24.40\pm0.17$ $^v$& $24.67\pm0.13$ $^t$ & $-3.44$ & 0.158\\
           & & $766.33^{+32.43}_{-31.11}$& & $758.58\pm61.77$ & $859.01\pm53.00$ &&\\
And\,XXII  & $21.07\pm0.08$ & $24.38\pm0.08$ & $21.11\pm0.07$ &$24.65\pm0.07$ $^u$ & $24.87\pm0.09$ $^u$  & $-3.44$ & 0.133\\
           & & $752.54^{+28.24}_{-27.24}$& &$853\pm26$ & $940.12\pm41.70$  &&\\
\hline
\end{tabular}
\tablecomments{
\footnotesize{TRGB$^a$ is the $i$-band tip magnitude of \citet{Conn2012}. $M_{\rm TRGB}$ is the adopted absolute tip magnitude in Eq.~(\ref{eq:mu}). Uncertainties on $\mu_{\,this\,work}$ include only the tip measurement. Literature values are from:
a \citep{Conn2012}, 
b \citep{2017ApJ...850..137M} , 
c \citep{1993AJ....106.1819K}, 
d \citep{2013MNRAS.435.3206D}, 
e \citep{2000ApJ...529..745F}, 
f \citep{2014ApJ...789..147W}, 
g \citep{1999AstL...25..332T}, 
h \citep{2009AJ....138..332J}, 
i \citep{2004ApJ...612L.121Z}, 
j \citep{2011ApJ...729...23B}, 
k \citep{2007ApJ...659L..21Z}, 
l \citep{2006MNRAS.371.1983M}, 
m \citep{2010MNRAS.407.2411C}, 
n \citep{2007ApJ...670L...9M}, 
o \citep{2017ApJ...837..102S}, 
p \citep{2016ApJ...819..147M}, 
q \citep{2008ApJ...676L..17I}, 
r \citep{2008ApJ...688.1009M}, 
s \citep{2013ApJ...779....7C}, 
t \citep{2009ApJ...705..758M}, 
u \citep{2013MNRAS.430...37C},
and v \citep{2015ApJ...806..200C}.}\\
}
\footnotesize{Galaxies marked with a star were previously studied by the INT group: And\,I \citep{2021ApJ...923..164S}, And\,VII \citep{2021ApJ...910..127N}, and And\,IX \citep{Abdollahi2023}. $A_i$ was computed from $E(B-V)$ of \citet{Schlegel1998} with the coefficient of \citet{Schlafly2011}.}
\end{table*}


\section{Results}
\label{sec:results}

Table~\ref{table:TRGB} lists the TRGB magnitudes and distances of the 17 satellites measured here, together with And~I, And~VII and And~IX from earlier survey papers (marked with a star). The distance moduli range from $23.38 \pm 0.06$\,mag (And~XVI, 474\,kpc) to $25.35 \pm 0.06$\,mag (And~XVIII, 1177\,kpc), and 15 of the 20 galaxies lie between 650 and 850\,kpc, around the distance of M31.

Figure~\ref{fig:distances} compares our distances with published values. For most satellites, our distance lies within the published range or agrees with at least one earlier estimate. And~XVIII falls just below the range, but agrees with \citet{Conn2012} within the uncertainties. The earlier values come from several methods (TRGB, RR~Lyrae, horizontal branch and CMD fitting) and telescopes, so this agreement is a useful check on our single, uniform analysis.

The closest comparison is with \citet{Conn2012}, who also measured the tip in the $i$ band. For 12 of the 18 galaxies in common, our tip is brighter by up to 0.39\,mag (And~XI). Six of these differences lie within $1\sigma$ and ten within $2\sigma$; only And~II (0.29\,mag) and And~XVI (0.22\,mag) exceed $2\sigma$. In the opposite direction, our tip for And~IX is 0.59\,mag fainter ($1.7\sigma$). Crowding may explain part of the offset: two red giants just below the tip that blend into one source look brighter than the tip and pull the edge to brighter magnitudes \citep{Madore2023}. Our INT images (0.33\,arcsec pixels) can be more affected by blending in the galaxy centres than the PAndAS CFHT/MegaCam data used by \citet{Conn2012}, which have finer pixels and cover each galaxy fully. The different methods, a Bayesian fit in \citet{Conn2012} and a Sobel filter here, may also play a role.

\section{Discussion and conclusions}
\label{sec:conclusions}

We have presented TRGB distances to 17 satellites of M31 from the $i$-band photometry of the INT monitoring survey, and combined them with earlier survey distances for three more. Most are consistent with at least one published estimate, and the common data set and analysis give a homogeneous basis for comparing these galaxies. Our tips tend to be brighter than those of \citet{Conn2012}; crowding and the different methods may contribute, but the comparison alone does not establish the cause.

The revised distances have particular implications for three systems. And~XII lies about 60\,kpc behind M31 (for an M31 distance of $\sim780$\,kpc), which is only about $1.5\sigma$ given our uncertainty; the earlier value of \citet{Conn2012} placed it farther behind. Our result is therefore consistent with the first-infall interpretation discussed by \citet{2007ApJ...662L..79C}, but does not add strong new support to it.

And~XXII is closer than all previous estimates. Its projected separation from M33 is 42\,kpc \citep{2009ApJ...705..758M}. Adopting the M33 distance of $794\pm23$\,kpc used in recent M31--M33 orbit models \citep[as summarised by][]{2022ApJ...938..101S}, our distance gives a three-dimensional separation of about 59\,kpc, compared with about 72\,kpc for the distance of 853\,kpc from \citet{2013MNRAS.430...37C}. Our result therefore keeps And~XXII close to M33. However, \citet{2022ApJ...938..101S} found that the M33 distance differs significantly from this value, so this separation depends on the adopted M33 distance, and the association cannot be established from distance alone \citep{2013MNRAS.430...37C}.

And~XVIII is closer than all previous estimates, but it still lies about 400\,kpc behind M31. Its membership in the M31 system therefore remains uncertain. Its distance at discovery was even larger \citep{2008ApJ...688.1009M}.

Together with the census of approximately 2800 LPV candidates across 17 M31 satellites, these distances contribute to a homogeneous characterization of the stellar populations in the Andromeda satellite system. Further constraints on membership will require the distances to be combined with kinematics and three-dimensional positions.

\scriptsize
\bibliographystyle{ComBAO}
\bibliography{ComBAO}

\end{document}